# Phase stability and physical properties of $Ta_5Si_3$ compounds from first-principles calculations

Xiaoma Tao, Philippe Jund, Catherine Colinet and Jean-Claude Tedenac

Institut Charles Gerhardt, Universite Montpellier 2, Pl. E. Bataillon CC1503
34095 Montpellier, France

Abstract: We present a study of the thermodynamic and physical properties of $Ta_5Si_3$ compounds by means of density functional theory based calculations. Among the three different structures ($D8_m$, $D8_l$, $D8_8$), the $D8_l$ structure ($Cr_5B_3$-prototype) is the low temperature phase with a high formation enthalpy of -449.20kJ/mol, the $D8_m$ structure ($W_5Si_3$-prototype) is the high temperature phase with a formation enthalpy of -419.36kJ/mol, and the $D8_8$ structure ($Mn_5Si_3$-prototype) is a metastable phase. The optimized lattice constants of the different $Ta_5Si_3$ compounds are also in good agreement with the experimental data. The electronic density of states (DOS) and the bonding charge density have also been calculated to elucidate the bonding mechanism in these compounds and the results indicate that bonding is mostly of covalent nature. The elastic constants of the $D8_m$ and $D8_l$ structures have been calculated together with the different moduli. Finally, by using a quasiharmonic Debye model, the Debye temperature, the heat capacity, the coefficient of thermal expansion and the Grüneisen parameter have also been obtained in the present work. The transformation temperature (2303.7K) between the $D8_m$ and the $D8_l$ structures has been predicted by means of the Gibbs energy, and this predicted temperature (2303.7K) is close to the experimental value (2433.5K).

## 1. Introduction

The transition metal silicides are of great interest as ultrahigh temperature materials, that have a variety of unique physical and chemical properties, such as good stability, high melting points, low densities, high resistance to oxidation, good creep tolerance and excellent mechanical strength at elevated temperature [1, 2, 3]. This has been illustrated recently by studies on $Mo_5Si_3$ based alloys [4, 5, 6].

As a member of refractory metals, tantalum is a candidate material for applications in high temperature hardware because of its high melting temperature and excellent ductility [7]. Usually the Ta-Si compounds are prepared by self-propagating high-temperature synthesis (SHS) from elemental powder compacts of corresponding stoichiometries [8, 9]. From the Ta-Si binary phase diagram [10], the $Ta_5Si_3$ phase has three different structures: $D8_m$, $D8_l$, and $D8_8$. The $D8_l$ $Cr_5B_3$-prototype structure is a low temperature phase, the $D8_m$ $W_5Si_3$-prototype structure is a high temperature phase and the $D8_8$ $Mn_5Si_3$ prototype structure is a metastable phase. The corresponding structural data [11] are summarized in Table 1. As a potential high temperature compound, the structural, electronic and elastic properties of $Ta_5Si_3$ are very important and essential for the material design and the material development. Unfortunately, there are few numerical studies focused on the electronic and mechanical properties of $Ta_5Si_3$. In order to lift this lack of data and to confront numerical results with the available experimental data, the density functional theory (DFT) and the quasiharmonic Debye model have been employed to study the physical and chemical properties of $Ta_5Si_3$.

The remainder of this paper is organized as follows. In Section 2, the method and calculation details are described. In section 3, the phase stability, electronic, structural and the elastic constants as well as some thermodynamic properties are presented and discussed. Finally, some conclusions are drawn in Section 4.

## 2. Computational details and theoretical background

First-principles calculations are performed by using the scalar relativistic all-electron Blöchl's projector augmented-wave (PAW) method [12,13] within the generalized gradient approximation (GGA), as implemented in the highly-efficient

Vienna ab initio simulation package (VASP) [14,15]. In the standard mode, VASP performs a fully relativistic calculation for the core-electrons and treats valence electrons in a scalar relativistic approximation. For the GGA exchange- correlation function, the Perdew-Wang parameterization (PW91) [16, 17] is employed. Here we adopted the standard version of the PAW potentials for Si and Ta atoms. For Si 4 electronic states are included in the valence shell whereas 5 electronic states are taken into account for Ta. A plane-wave energy cutoff of 450 eV is held constant for the study of the different $Ta_5Si_3$ compounds. Brillouin zone integrations are performed using Monkhorst-Pack K-point meshes [18], and the Methfessel-Paxton technique [19] with a smearing parameter of 0.2 eV. The reciprocal space (k-point) meshes are increased to achieve convergence to a precision of 1meV/atom. The total energy is converged numerically to less than $1\times10^{-6}$ eV/unit with respect to electronic, ionic and unit cell degrees of freedom, and the latter two are relaxed using calculated forces with a preconditioned conjugated gradient algorithm. After structural optimization, calculated forces are converged to less than 0.01eV/Å. All calculations are performed using the "high" precision setting within the VASP input file to avoid wrap-around errors.

Initially, the three structures have been optimized by fully relaxing the simulation cell (ionic positions, shape, and volume) starting from the experimental parameters taken from the Pearson's Handbook [11]. Then, in order to obtain the calculated equilibrium characteristics of each structure, several total energy calculations have been performed at different fixed volumes while the ionic positions and the cell shape could vary. These several total energies obtained for several different volumes were then fitted with the Vinet [20] equation of state. This permits to calculate the equilibrium volume ($\Omega_0$), the total energy (E), the bulk modulus (B) and the pressure derivative of the bulk modulus ($\partial B/\partial P$).

The formation enthalpy of the $Ta_5Si_3$ alloys can be calculated from the following equation:

$$\Delta H(Ta_5Si_3) = E(Ta_5Si_3) - 5E(Ta) - 3E(Si) \tag{1}$$

where $E(Ta_5Si_3)$, $E(Ta)$ and $E(Si)$ are the equilibrium first-principles calculated total energies of the $Ta_5Si_3$ compound, Ta with bcc structure and Si with diamond structure, respectively.

The elastic constants can be obtained by calculating the total energy as a function of the strains. For a small strain (ε') on a solid, the Hook's law is applicable and the elastic energy $\Delta E$ is a quadratic function of the strains

$$\Delta E = V \sum_{i,j=1}^{6} \frac{1}{2} C_{ij} e_i e_j \tag{2}$$

where $C_{ij}$ are the elastic constants, $V$ is the total volume of the unit cell, and $e_i$ are the components of the strain matrix [21]:

$$\varepsilon' = \begin{pmatrix} e_1 & e_6 & e_5 \\ e_6 & e_2 & e_4 \\ e_5 & e_4 & e_3 \end{pmatrix} \tag{3}$$

Under these strains, the original lattice vectors ***a*** would be transformed into the strained ones ***a'*** as ***a***'=(***I***+$\varepsilon'$)***a***, where ***I*** is the identity matrix.

For a cubic crystal, there are only three independent elastic constants, namely, $C_{11}$, $C_{12}$ and $C_{44}$. Furthermore, the bulk modulus $B$ and the tetragonal shear constant $C'$ are related to $C_{11}$ and $C_{12}$ by: $B= (C_{11} +2C_{12})/3$, and $C'= (C_{11}-C_{12})/2$. $C_{44}$ is the trigonal shear constant. The bulk modulus $B=V(d^2E/dV^2)$ is determined by fitting an equation of state [20] with the calculated total energy $E$ as a function of volume $V$. For the calculation of the tetragonal shear modulus, $C_{11}$-$C_{12}$ was determined by using a volume-conserving tetragonal strain

$$\varepsilon' = \begin{pmatrix} e_1 & 0 & 0 \\ 0 & e_1 & 0 \\ 0 & 0 & (1+e_1)^{1/2} - 1 \end{pmatrix} \tag{4}$$

with energy change

$$\Delta E(e_1) = V(C_{11} - C_{12}) e_1^2 + O(e_1^4) \tag{5}$$

The remaining independent modulus, $C_{44}$, was found by shearing the crystal with monoclinic strain

$$\varepsilon' = \begin{pmatrix} 0 & \frac{1}{2}e_6 & 0 \\ \frac{1}{2}e_6 & 0 & 0 \\ 0 & 0 & e_6^2/\left(4-e_6^2\right) \end{pmatrix} \quad (6)$$

The associated energy change is

$$\Delta E(e_6) = \frac{1}{2}VC_{44}e_6^2 + O\left(e_6^4\right) \quad (7)$$

For a tetragonal material, there are six independent single-crystal elastic constants, that is, $C_{11}$, $C_{12}$, $C_{33}$, $C_{13}$, $C_{44}$ and $C_{66}$. The details of the calculations can be found in Ref. [6] and are not recalled here.

The effective elastic moduli of polycrystalline aggregates are usually calculated by two approximations due to Voigt [22] and Reuss [23] in which respectively uniform strain or stress are assumed throughout the polycrystal. Hill [24] has shown that the Voigt and Reuss averages are limits and suggested that the actual effective moduli can be approximated by the arithmetic mean of the two bounds, referred to as the Voigt–Reuss–Hill (VRH) value from the reference [25].

The Voigt bounds for cubic systems are

$$B_V = \tfrac{1}{3}\left(C_{11} + 2C_{12}\right) \quad (8)$$

$$G_V = \tfrac{1}{5}\left(C_{11} - C_{12} + 3C_{44}\right) \quad (9)$$

and the Reuss bounds are

$$B_R = \tfrac{1}{3}\left(C_{11} + 2C_{12}\right) \quad (10)$$

$$G_R = \frac{5\left(C_{11} - C_{12}\right)C_{44}}{4C_{44} + 3\left(C_{11} - C_{12}\right)} \quad (11)$$

The Voigt bounds for tetragonal systems are

$$B_V = \tfrac{1}{9}\left[2\left(C_{11} + C_{12}\right) + C_{33} + 4C_{13}\right] \quad (12)$$

$$G_V = \tfrac{1}{30}\left(M + 3C_{11} - 3C_{12} + 12C_{44} + 6C_{66}\right) \quad (13)$$

and the Reuss bounds are

$$B_R = C^2/M \quad (14)$$

$$G_R = 15\left\{\left(18B_V/C^2\right) + [6/\left(C_{11} - C_{12}\right)] + \left(6/C_{44}\right) + \left(3/C_{66}\right)\right\}^{-1} \quad (15)$$

$$M = C_{11} + C_{12} + 2C_{33} - 4C_{13} \quad (16)$$

$$C^2 = (C_{11} + C_{12})C_{33} - 2C_{13}^2 \tag{17}$$

Finally, the VRH mean values are obtained by

$$B = \frac{1}{2}(B_V + B_R) \tag{18}$$

$$G = \frac{1}{2}(G_V + G_R) \tag{19}$$

The Young's modulus and the Poisson's ratio can be obtained from the bulk and shear moduli [25]

$$E = \frac{9BG}{3B + G} \tag{20}$$

$$v = \frac{3B - 2G}{2(3B + G)} \tag{21}$$

To investigate the thermodynamic properties of the $Ta_5Si_3$, we apply here the quasiharmonic Debye model [26], in which the non-equilibrium Gibbs function $G^*$(V; P, T) takes the form of [26]

$$G^*(V;P,T) = E(V) + PV + A_{Vib}(\Theta(V);T) \tag{22}$$

where *E(V)* is the total energy per unit, *PV* corresponds to the constant hydrostatic condition, $\Theta(V)$ is the Debye temperature, and vibrational term $A_{Vib}$ can be expressed as [27,28]

$$A_{Vib}(\Theta;T) = nk_BT\left[\frac{9}{8}\frac{\Theta}{T} + 3\ln(1 - e^{-\Theta/T}) - D(\Theta/T)\right] \tag{23}$$

where $D(\Theta/T)$ represents the Debye integral, *n* is the number of atoms per formula unit, $k_B$ is Boltzmann constant and *T* is the absolute temperature.

The Debye temperature can be estimated from the average sound velocity using the following equation:

$$\Theta = \frac{h}{k_B}\left(\frac{3}{4\pi V}\right)^{1/3} v_m \tag{24}$$

where *h* is the Plank's constant and V is the atomic volume. The average sound velocity in a polycrystalline system, $v_m$ is evaluated by

$$v_m = \left[\frac{1}{3}\left(\frac{2}{v_t^3} + \frac{1}{v_l^3}\right)\right]^{-1/3} \tag{25}$$

where $v_t$ and $v_l$ are the mean transverse and longitudinal sound velocities, which can be related to the shear and bulk moduli by the Navier's equations:

$$v_l = \left(\frac{3B + 4G}{3\rho}\right)^{1/2} \text{ and } v_l = \left(\frac{G}{\rho}\right)^{1/2} \tag{26}$$

Finally, $\Theta$ is expressed as [27]

$$\Theta = \frac{\hbar}{k_B}\left[6\pi^2 V^{1/2} n\right]^{1/3} f(v)\sqrt{\frac{B_S}{M}} \tag{27}$$

where $M$ is the molecular mass per unit cell, $B_S$ the adiabatic bulk modulus, which can be approximated by the static compressibility [26]

$$B_S \cong B(V) = V\left(\frac{d^2E(V)}{dV^2}\right) \tag{28}$$

the Poisson's ratio $v$ is taken from Eq.(21), $f(v)$ is given by Refs. [28,29]. Therefore, the non-equilibrium Gibbs function $G^*(V;P,T)$ as a function of $V$, $P$ and $T$ can be minimized with respect to the volume $V$

$$\left(\frac{\partial G^*(V;P,T)}{\partial V}\right)_{P,T} = 0 \tag{29}$$

By solving Eq. (29), we get the thermal equation of state. The isothermal bulk modulus $B_T$ and the heat capacity $C_V$ are given by [30]

$$B_T(P,T) = V\left(\frac{\partial^2 G^*(V;P,T)}{\partial V^2}\right)_{P,T} \tag{30}$$

$$C_V = 3nk_B\left[4D(\Theta/T) - \frac{3\Theta/T}{e^{\Theta/T} - 1}\right] \tag{31}$$

Anharmonic effects of the vibrating lattice are usually described in terms of a Grüneisen parameter $\gamma$, which can be defined as

$$\gamma = \frac{\partial \ln \Theta_D}{\partial \ln V} \tag{32}$$

$$\alpha = \frac{\gamma C_V}{B_T V} \tag{33}$$

where $\alpha$ is the thermal expansion coefficient.

## 3. Results and discussion

### 3.1 Phase stability

As shown in eq. 1, in the present work, we have used the silicon diamond and the tantalum bcc phases as reference states. The calculated lattice constants and formation enthalpies have been reported in Table 2 together with available experimental data and other theoretical data. The lattice constants of the pure elements are in good agreement with the literature.

For the $Ta_5Si_3$ system, the present optimized structures (at 0K) are also in good agreement with the experimental lattice constants [31-36]. Discrepancies of lattice constants for the three prototype structures of $Ta_5Si_3$ are less than 1% in comparison to the experimental data [31-36]. However the present calculated lattice constants are slightly larger than the experimental data, which is a common error inherent to GGA calculations.

The calculated formation enthalpies of the three different structures obtained from the plot of the total energies vs atomic volume (Fig.1) and the subsequent fit, indicate that the $Cr_5B_3$-prototype phase is the stable phase with the lowest energy. Among the three phases, the formation enthalpy of the $W_5Si_3$-prototype phase is higher than that of the $Cr_5B_3$-prototype phase by 29.84 kJ/mol, and the formation enthalpy of the $Mn_5Si_3$-prototype phase is higher than that of the $Cr_5B_3$-prototype phase by 82.80 kJ/mol. All of these results coincide with the experimental information that the $Cr_5B_3$-prototype phase is a low temperature phase, the $W_5Si_3$-prototype phase is a high temperature phase and the $Mn_5Si_3$-prototype phase is a metastable phase [10].

Compared with the available experimental data [37-40] (see Table2), the present calculated value of the enthalpy of formation of $Ta_5Si_3$ ($Cr_5B_3$-prototype structure) is in good agreement with the values reported by Schlesinger [10]. These latter values have been obtained from the experimental data of Myers and Searcy [39] using a Knudsen effusion method and Levine and Kolodney [40] using electromotive force measurements. However our calculated value is strongly more negative than the ones obtained by Robins and Jenkins [37] and by Meschel and Kleppa [38]. In these two works [37, 38], calorimetric methods were used. Meschel and Kleppa [38] also determined the enthalpy of formation of the $TaSi_2$ compound and obtained a value of -368.00 kJ/mol, which is more negative than the value they obtained for $Cr_5B_3$-prototype $Ta_5Si_3$ (-304.80 kJ/mol), this appears very unlikely; indeed the experimental Ta-Si phase diagram [10] indicates a melting temperature of $Ta_5Si_3$ (2550°C) much higher than the one of $TaSi_2$ (2040°C). Meschel and Klepppa [41] also performed experimental determinations of enthalpies of formation of intermetallic compounds in the V-Si and Nb-Si systems. In both cases, they obtained more negative values for the enthalpy of formation of $A_5Si_3$ (A=V, Nb) compound than for the one of $ASi_2$ compound. In the V-Si systems, this point is confirmed by recent ab-initio calculations [42,43].

## 3.2 Electronic structure

The total (TDOS) and partial (PDOS) density of states for the $Cr_5B_3$-prototype, $W_5Si_3$-prototype and $Mn_5Si_3$-prototype structures are shown in Fig.2 and in Fig.3. In Fig.2 (a), the states, which are approximately located between -5eV and -1eV below the Fermi level, originate from the bonding of Si-p and Ta-*d* states. The Fermi level lies about 0.5 eV above the pseudogap minimum, i.e. the bonding states are completely occupied, and the TDOS (Fig.2(a)) at the Fermi level is rather low (0.358 states/eV/atom). The bonding states, as seen in Fig.3(a), are dominated by Si-*s*, Si-*p* and Ta-*d*, while the antibonding states are dominated by Ta-*d* states. Below the Fermi level, Si-*p* and Ta-*d* peaks show evidence for strong hybridization, while the other contributions are small. For the $W_5Si_3$-prototype phase, it is found that the main bonding peaks between -11.5 eV and -8 eV are predominantly derived from Si-s states, the main bonding peaks between -6.5 eV and -2eV are dominated by Si-p and Ta-d states, while the main bonding peaks from -2 eV to 2.5 eV originate mainly from Ta-d states. Si-p and Ta-d states show evidence for hybridization below the Fermi level. The $Mn_5Si_3$-prototype phase has PDOS (Fig.3(b)) similar to the ones of the $W_5Si_3$-prototype phase. The three TDOS have some similarities, however in Fig.2 the TDOS at the Fermi level are different, the $Cr_5B_3$-prototype phase has the smallest $n(E_f)$ with 0.358 states/eV/atom, the $n(E_f)$ of the $W_5Si_3$-prototype phase is equal to 0.474 states/eV/atom, and the $Mn_5Si_3$-prototype phase has the largest $n(E_f)$ with 0.697 states/eV/atom. Generally speaking, the smaller $n(E_f)$ is, the more stable the compound is. So in agreement with the results obtained from the formation enthalpies and in agreement with the experimental data [10], the most stable phase is the $Cr_5B_3$ phase followed by the $W_5Si_3$ phase and finally the $Mn_5Si_3$ phase.

To visualize the nature of the bonds and to explain the charge transfer and the bonding properties of $Ta_5Si_3$, we have investigated the bonding charge density. The bonding charge density, also called the deformation charge density, is defined as the difference between the self-consistent charge density of the interacting atoms in the compound and a reference charge density constructed from the super-position of the non-interacting atomic charge density at the crystal sites. The distribution of the bonding charge densities in the (001), (100) and (110) planes of the $Cr_5B_3$-prototype phase have been calculated to understand the alloying mechanism and are shown in Fig. 4 (a), (b) and (c) , respectively.

In Fig. 4, the bonding charge density shows a depletion of the electronic density at the lattice sites together with an increase of the electronic density in the interstitial region. In Fig. 4(a), the covalent character of the bonding is dominant in the (001) plane and the build-up of bonding charge along the Ta-Si bond direction is very strong. This feature is consistent with the PDOS plots, in Fig. 3 (a), showing the importance of the Ta-d and Si-p hybridization, associated with the Ta-Si bonding. A significant redistribution of the bonding charge in the interstitial region is also seen in Fig. 4(b) and (c). The bonding between Ta and Si in the [100] direction is not so strong than the bonding between Ta and Ta in the [110] direction. From Fig.4 (c), the Si-Si bonding in the [110] direction is also stronger than the Ta-Si and the Ta-Ta bonding.

### 3.3 Elastic properties

In general, elastic properties of a solid are very important because they permit to obtain the mechanical properties such as the bulk modulus, the shear modulus, the Young's modulus and the Poisson's ratio. The strength of a material or the sound velocity can be obtained from the elastic constants as well. In order to shed some light on the mechanical properties of $Ta_5Si_3$ compounds, the elastic constants of the pure elements, of the $Cr_5B_3$-prototype phase and of the $W_5Si_3$-prototype phase have been calculated in the present work and are listed in Table 3. Concerning the calculated elastic constants of the two pure elements Ta and Si, the present results are in good agreement with the experimental data [44, 45]. The requirement of mechanical stability in a cubic crystal leads to the following restrictions on the elastic constants, $C_{11}$- $C_{12}$ > 0, $C_{11}$ > 0, $C_{44}$ > 0, $C_{11}$ + $2C_{12}$ > 0. Obviously, the calculated elastic constants of the pure elements are satisfying those mechanical stability conditions.

For the tetragonal system, the mechanical stability criteria are given by $C_{11}>0$, $C_{33}>0$, $C_{44}>0$, $C_{66}>0$, $(C_{11}-C_{12})>0$, $(C_{11}+C_{33}-2C_{13})>0$, and $[2(C_{11}+C_{12})+C_{33}+4C_{13}]>0$. The calculated elastic constants of the $Cr_5B_3$-prototype phase and the $W_5Si_3$-prototype phase satisfy these stability conditions. In the present calculations, $C_{11}>C_{33}$, indicating that the bonding strength along the [100] and [010] direction is stronger than that of the bonding along the [001] direction. This is consisted with the analysis of the bonding charge density (Fig.4). It is worth noting that the $C_{44}$ and $C_{66}$ relationship of the $Cr_5B_3$-prototype phase is different from that of the $W_5Si_3$-prototype phase. For the $Cr_5B_3$-prototype phase, $C_{44}>C_{66}$ whereas for the $W_5Si_3$-prototype phase, $C_{44}<C_{66}$. This suggests that the [100](010) shear is easier than the [100](001) shear for the $Cr_5B_3$-prototype phase, whereas the contrary is true for the $W_5Si_3$-prototype phase. The shear elastic anisotropy factor [46] can be estimated as $A=2C_{66}/(C_{11}-C_{12})$. If A is equal to one, no anisotropy exists. A=0.835 for the $Cr_5B_3$-prototype phase and A=1.005 for the $W_5Si_3$-prototype phase, indicating that the shear elastic anisotropy of the $Cr_5B_3$-prototype phase is larger than that of the $W_5Si_3$-prototype phase. The shear elastic anisotropy factor of the $W_5Si_3$-prototype phase is very close to 1, which indicates that the shear elastic properties of the (001) plane are nearly independent of the shear direction. Similar properties have been found for $Mo_5Si_3$ [6, 47].

The bulk modulus, shear modulus, Young's modulus and Poisson's ratio have been estimated from the calculated single crystal elastic constants, and are given in Table 3. It can be seen that the bulk moduli of the $Cr_5B_3$-prototype phase and the $W_5Si_3$-prototype phase are slightly larger than that of pure Ta. However, the shear and Young's moduli of $Ta_5Si_3$ with the $Cr_5B_3$-prototype structure and the $W_5Si_3$-prototype structure are two times larger than the shear modulus of Ta. The present calculated bulk moduli, shear moduli, Young's moduli and Poisson's ratio of $Ta_5Si_3$ with the $Cr_5B_3$-prototype and the $W_5Si_3$-prototype structure are close to those of $Mo_5Si_3$ with the $W_5Si_3$-prototype structure [47] even though the moduli of Ta are almost half of those of Mo indicating that the bonding in $Ta_5Si_3$ is stronger than the one in $Mo_5Si_3$.

The ratio between the bulk and the shear modulus, B/G, has been proposed by Pugh [48] to predict brittle or ductile behavior of materials. According to the Pugh criterion, a high B/G value indicates a tendency for ductility. If B/G>1.75, then ductile behavior is predicted, otherwise the material behaves in a brittle manner. The ratio for the $Cr_5B_3$-prototype phase is close to 1.75, while it is larger than 1.75 for the $W_5Si_3$-prototype phase. These results suggest that the $Cr_5B_3$-prototype phase is slightly prone to brittleness, and the $W_5Si_3$-prototype phase is more prone to ductility.

### 3.4 Thermodynamic properties

Through the quasiharmonic Debye model, one can calculate the thermodynamic quantities of $Ta_5Si_3$ at any temperature and pressure, from the E-V data calculated at T=0 and P=0. In the present work, the thermal properties are determined in the temperature range from 0 to 3000K, and the pressure effect is studied in the 0 to 20 GPa range. The temperature effects on the volumes of the $Cr_5B_3$-type structure are shown in Fig.5 (in this section similar results are observed for the $W_5Si_3$-type structure and are thus not shown). As expected the volume increases with increasing temperature, and the rate of increase is high. In Fig.6 we report the evolution of the bulk modulus as a function of T at different pressures for the $Cr_5B_3$-type structure. It is worth noting from the regular spacing of the curves observed in Fig. 6 that the relationship between the bulk modulus of the $Cr_5B_3$-type phase and the pressure is nearly linear at various temperatures ranging from 0 to 3000K. The bulk modulus increases with pressure at a given temperature and decreases with temperature at given pressure. These results are due to the fact that the effect of increasing the pressure on the material is the same as decreasing the temperature of the material.

The knowledge of the heat capacity of a crystal not only provides essential informations on its vibrational properties but is also mandatory for many applications.

Fig. 7 and Fig. 8 represent the variation of the heat capacity, $C_V$(T), and volume expansion coefficient, α(T) as a function of temperature, respectively. These two quantities show a sharp increase up to ~400K, and at high temperature $C_V$ is close to a constant, which is the so-called Dulong-Petit limit [21]. In Fig. 8, we have plotted the thermal expansion coefficient α of the $Cr_5B_3$-type structure. It is shown that, for a given pressure, α increases with temperature at low temperature especially at 0 pressure and gradually tends to a linear increase at high temperature. With increasing pressure, the increase of α with temperature becomes smaller. For a given temperature, α decreases strongly with increasing pressure, and remains very small at high temperatures and high pressures.

The calculated longitudinal, transverse and average sound velocities, Debye temperature and Grüneisen parameters of $Ta_5Si_3$ with the $Cr_5B_3$-prototype and the $W_5Si_3$-prototype structures have been calculated and listed in Table 4. The Debye temperatures of the $Ta_5Si_3$ with the $Cr_5B_3$-prototype and the $W_5Si_3$-prototype structures are larger than that of pure Ta (240 K) and smaller than that of Si (645K) [21], but are close to the average Debye temperature of $Ta_5Si_3$ (392K). The average sound velocity and Debye temperature of $Ta_5Si_3$ are smaller than the values obtained for $Mo_5Si_3$ [47]. As far as we know, there are no experimental values available for $Ta_5Si_3$ and we are not aware of other theoretical data for these quantities. Therefore our calculations should be taken as predictions and we hope that soon these predictions can be compared to either experimental or simulation data.

In order to elucidate the phase transformation between the $Cr_5B_3$-prototype and $W_5Si_3$-prototype structures at high temperature, the Gibbs energies of the two structures have been calculated at different temperatures and their difference is plotted in Fig. 9. From this figure, the phase transformation occurs at T=2303.7K, which is very close to the experimental temperature of 2433.5K [10]. This indicates the accuracy of the present calculations.

## 4. Conclusion

We have presented results of first-principles calculations for the phase stability, electronic structures, elastic properties and thermodynamic properties of $Ta_5Si_3$ compounds. The calculated results accurately predict the stable phase with the $Cr_5B_3$-type structure at low temperature and with the $W_5Si_3$-type structure at high temperature. The calculated lattice constants are in good agreement with experimental data. The density of states and bonding charge densities of $Ta_5Si_3$ have been calculated and indicate that covalent bonding is dominant in the (001) plane. The elastic constants and elastic moduli of $Ta_5Si_3$ have been predicted in this work, but as far as we know, there are no experimental data available for these quantities. Experiments are expected to validate the present numerical data. Using the Debye model, the heat capacity, thermal expansion, Debye temperature and Grüneisen parameters have been estimated. From the Gibbs energy of the $Cr_5B_3$-prototype and $W_5Si_3$-prototype phases, the temperature of phase transformation has been found at 2303.7 K, which is very close to the experimental value of 2433.5 K. The present results give hints for the design of novel materials based on $Ta_5Si_3$ compounds and should be used to stimulate future experimental and theoretical work.

Acknowledgment: X.T. thanks the “Fondation d’enterprise E.A.D.S.” for supporting his postdoctoral position grant.

**Table Captions**

Table 1 Crystallographic structural data for the three prototypical structures considered for the $Ta_5Si_3$ compounds [11]

Table 2 The calculated lattice constants and formation enthalpies for the $Ta_5Si_3$ compounds

Table 3 Calculated elastic constants, bulk modulus (B), shear modulus (G), Young's modulus (E) and Poisson's ratio ($v$) for the $Ta_5Si_3$ compounds

Table 4 Calculated longitudinal, transverse and average sound velocities (in m/s), Debye temperature (in Kelvin) and Grüneisen parameters of $Ta_5Si_3$ with the $Cr_5B_3$-prototype and $W_5Si_3$-prototype structures.

**Figure Captions**

Fig. 1 Total energy vs volume for the $Cr_5B_3$, $W_5Si_3$ and $Mn_5Si_3$-prototype structures.

Fig. 2. Total density of states (DOS) of the $Cr_5B_3$-prototype, $W_5Si_3$-prototype and $Mn_5Si_3$-prototype phases

Fig. 3 Partial density of states (DOS) of the $Cr_5B_3$-prototype, $W_5Si_3$-prototype and $Mn_5Si_3$-prototype phases

Fig.4 Bonding charge density in the (001), (100) and (110) planes of the $Cr_5B_3$-prototype

Fig. 5 Volume vs temperature at various pressures

Fig. 6 Relationship between the bulk modulus and the temperature at various pressures

Fig. 7 Relationship between the heat capacity and the temperature at various pressures

Fig. 8 Relationship between the thermal expansion and the temperature at various pressures

Fig. 9 The difference between the Gibbs energy of the $Cr_5B_3$-type structure and the $W_5Si_3$-type structure as a function of temperature

Table 1 Crystallographic structural data for the three prototypical structures considered for the $Ta_5Si_3$ compounds [11]

| prototype | Pearson symbol | Space group | Strukturbericht | | | Atomic Wyckoff positions |
|---|---|---|---|---|---|---|
| $Cr_5B_3$ | tI32 | I4/mcm (140) | $D8_1$ | Cr(1) | 16l | (x, x+1/2, z) |
| | | | | Cr(2) | 4c | (0, 0, 0) |
| | | | | B(1) | 4a | (0, 0, 1/4) |
| | | | | B(2) | 8h | (x, x+1/2, 0) |
| $Mn_5Si_3$ | hP16 | $P6_3$/mcm(193) | $D8_8$ | Mn(1) | 4d | (1/3, 2/3, 0) |
| | | | | Mn(2) | 6g | (x, 0, 1/4) |
| | | | | Si(1) | 6g | (x, 0, 1/4) |
| $W_5Si_3$ | tI32 | I4/mcm (140) | $D8_m$ | W(1) | 16k | (x, y, 0) |
| | | | | W(2) | 4b | (0, 1/2, 1/4) |
| | | | | Si(1) | 8h | (x, x+1/2, 0) |
| | | | | Si(2) | 4a | (0, 0, 1/4) |

Table 2 The calculated lattice constants and formation enthalpies for the $Ta_5Si_3$ compounds

| Phase | Prototype | Lattice parameters | | Internal parameters | | Formation enthalpy (kJ/mol) |
|---|---|---|---|---|---|---|
| | | a(Å) | c(Å) | Present | Exp. [11] | |
| Ta | W(A2) | 3.3069 | - | - | - | - |
| | | 3.303[10] | | | | |
| $Ta_5Si_3$ | $Cr_5B_3$ | 6.560 | 11.917 | $x_{Ta}$=0.1647 | $x_{Ta}$= 0.166 | -449.20 |
| | | 6.503 | 11.849[31] | $z_{Ta}$=0.1505 | $z_{Ta}$=0.15 | -419.20[10] |
| | | 6.504 | 11.849[32] | $x_{Si}$=0.3699 | $x_{Si}$=0.375 | -439.20[10] |
| | | 6.519 | 11.87[33] | | | -317.60[37] |
| | | 6.516 | 11.873[34] | | | -304.80[38] |
| | $Mn_5Si_3$ | 7.532 | 5.2578 | $x_{Ta}$=0.2457 | $x_{Ta}$=0.2361 | -366.40 |
| | | 7.459 | 5.215[35] | $x_{Si}$=0.6050 | $x_{Si}$=0.6001 | |
| | | 7.480 | 5.251[36] | | | |
| | | 7.469 | 5.212[32] | | | |
| | | 7.474 | 5.225[34] | | | |
| | $W_5Si_3$ | 10.01 | 5.1063 | $x_{Ta}$=0.0751 | $x_{Ta}$=0.074[a] | -419.36 |
| | | 9.86 | 5.05[32] | $y_{Ta}$=0.2204 | $y_{Ta}$=0.223[a] | |
| | | 9.892 | 5.042[33] | $x_{Si}$=0.1655 | $x_{Si}$=0.17[a] | |
| Si | Si(A4) | 5.4676 | - | - | - | - |
| | | 5.4306[10] | | | | |

a: this experimental data is taken from prototypical compound.

Table 3 The calculated elastic constants, bulk modulus (B), shear modulus (G), Young's modulus (E) and Poisson's ratio ($v$) for the $Ta_5Si_3$ compounds

| Structural | Ta (BCC) | Si(Diamond) | $Cr_5B_3$-type | $W_5Si_3$-type | $Mo_5Si_3$[a] |
|---|---|---|---|---|---|
| $C_{11}$ (GPa) | 284.24 | 153.34 | 412.65 | 410.19 | 446 |
| | 266[44] | 168[45] | | | |
| $C_{12}$ (GPa) | 158.85 | 56.09 | 116.65 | 145.66 | 174 |
| | 158[44] | 65[45] | | | |
| $C_{33}$ (GPa) | - | - | 361.91 | 338.89 | 390 |
| $C_{13}$ (GPa) | - | - | 144.81 | 125.45 | 140 |
| $C_{44}$ (GPa) | 66.83 | 75.19 | 135.95 | 92.94 | 110 |
| | 87[44] | 80[45] | | | |
| $C_{66}$ (GPa) | - | - | 123.61 | 132.91 | 140 |
| B(GPa) | 200.65 | 88.51 | 222.11 | 215.66 | 242 |
| | 194[44] | 99[45] | | | |
| G(GPa) | 65.14 | 63.13 | 130.57 | 112.98 | 126 |
| E(GPa) | 176.35 | 153.02 | 327.53 | 288.56 | 323 |
| $v$ | 0.35 | 0.21 | 0.254 | 0.277 | 0.278 |
| B/G | 3.08 | 1.40 | 1.70 | 1.91 | 1.92 |

[a] Ref. [47]

Table 4 The calculated longitudinal, transverse and average sound velocities (in m/s), Debye temperature (in Kelvin) and Grüneisen parameters of $Ta_5Si_3$ with the $Cr_5B_3$-prototype and $W_5Si_3$-prototype structures

| | $v_l$ | $v_t$ | $v_m$ | Θ | Γ |
|---|---|---|---|---|---|
| $Cr_5B_3$-type | 5562 | 3193 | 3546 | 414 | 2.0 |
| $W_5Si_3$-type | 5342 | 2967 | 3304 | 388 | 1.96 |
| $Mo_5Si_3$[47] | - | - | 4317 | 566 | - |

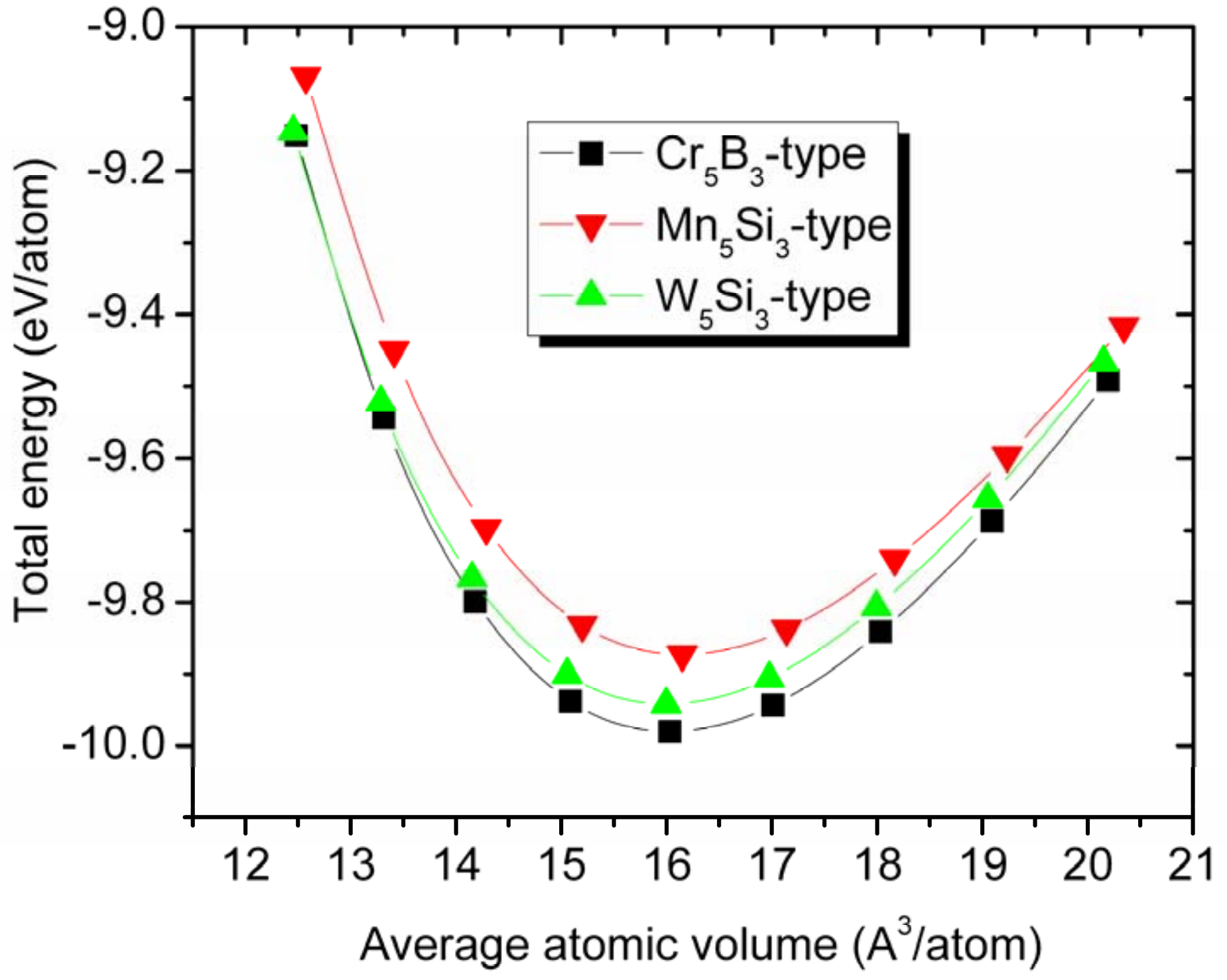

Figure 1

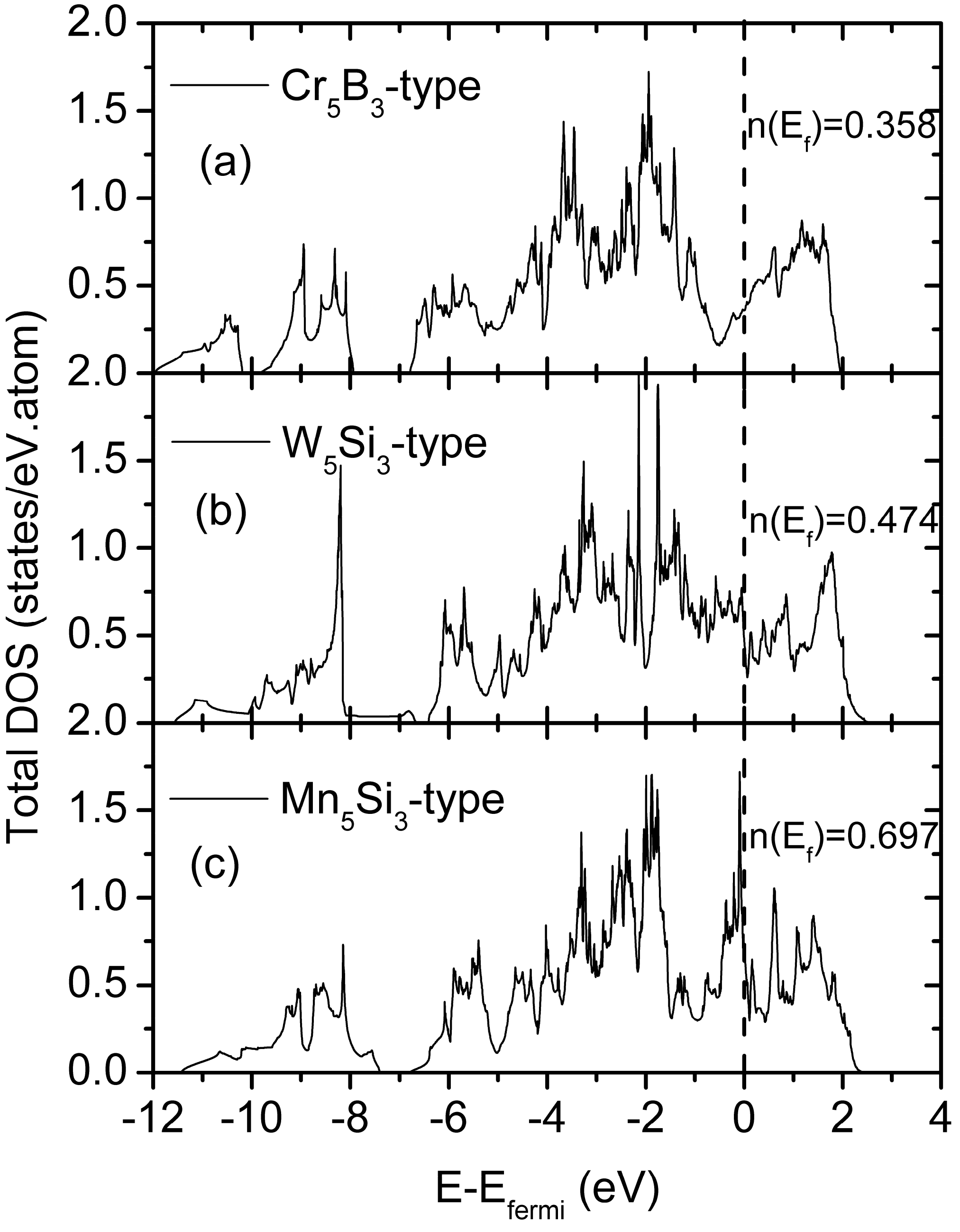

Figure 2

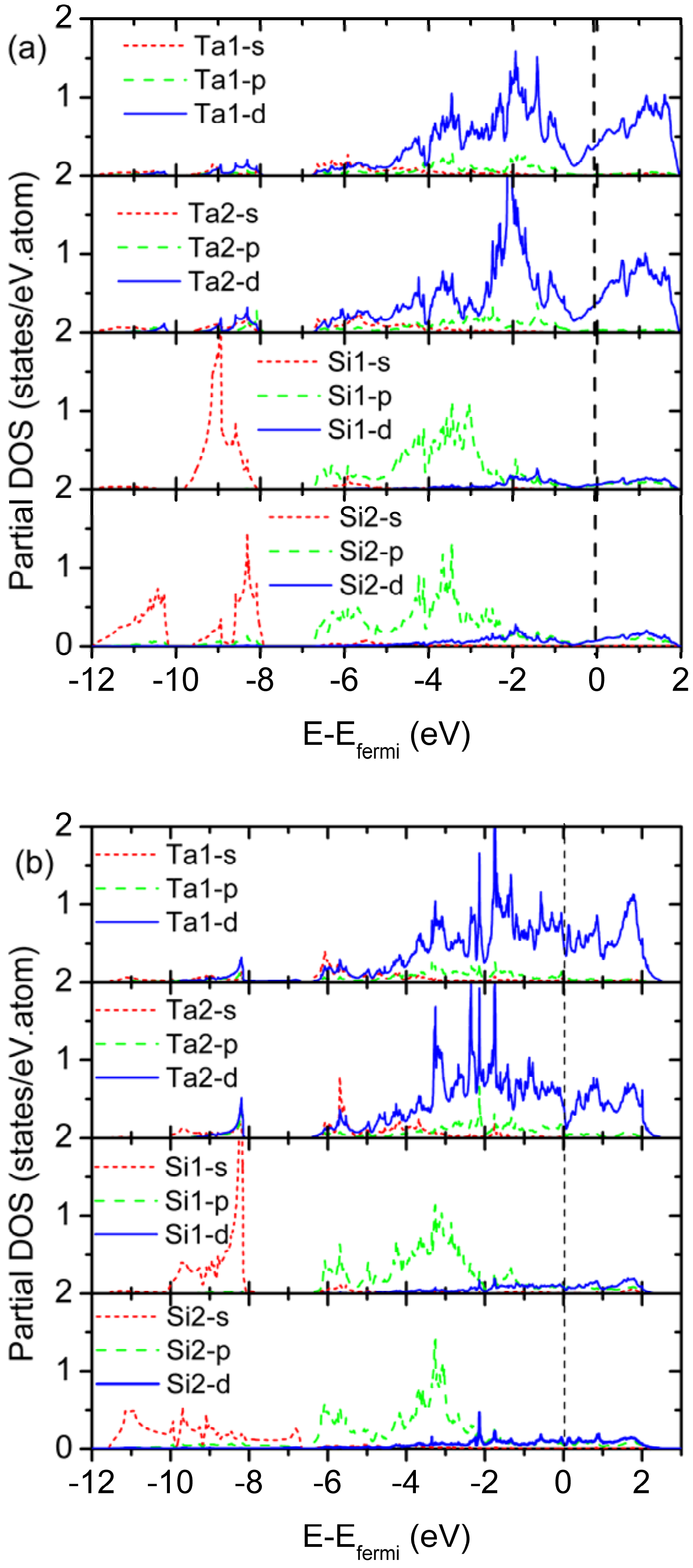

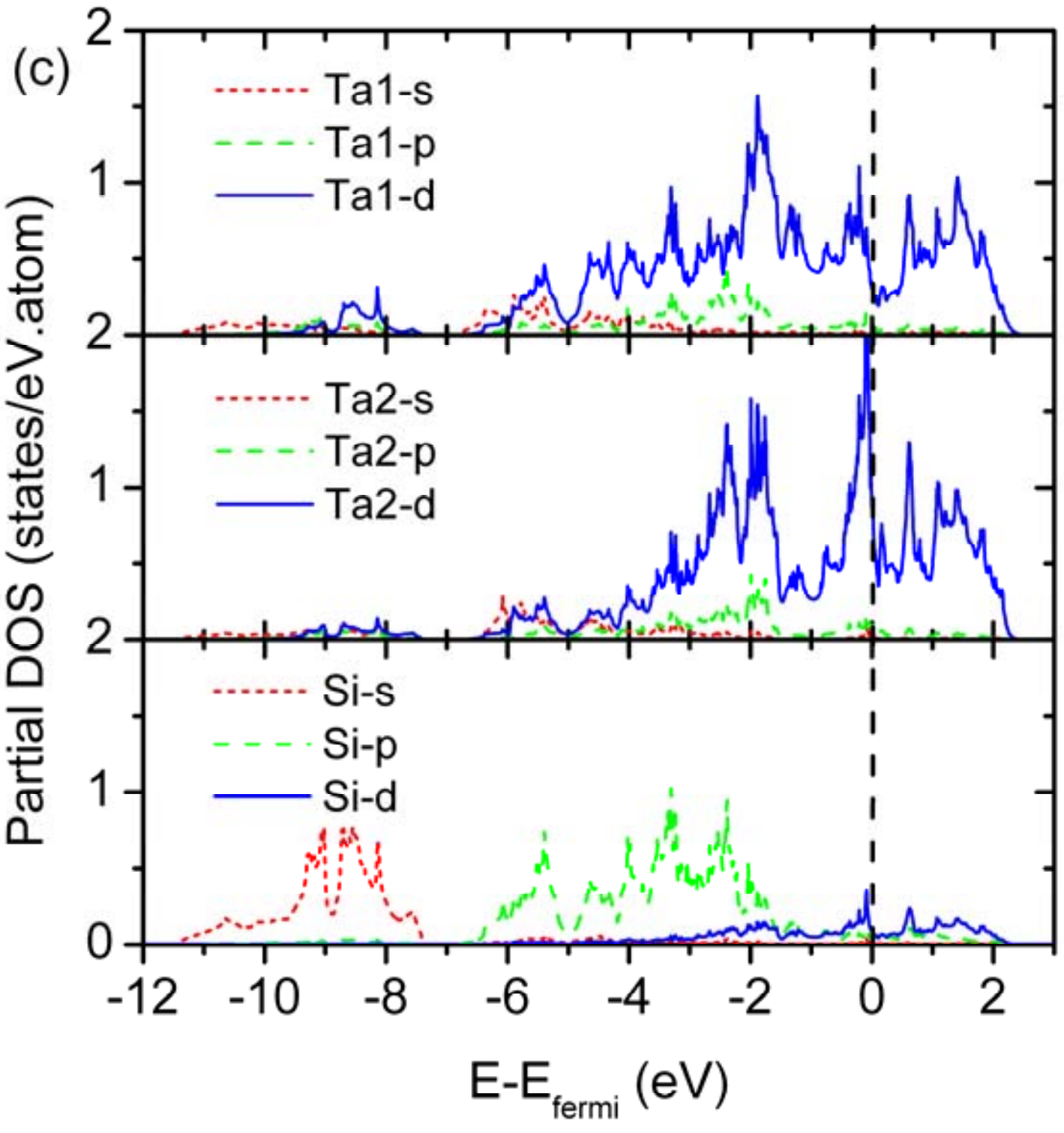

Figure 3

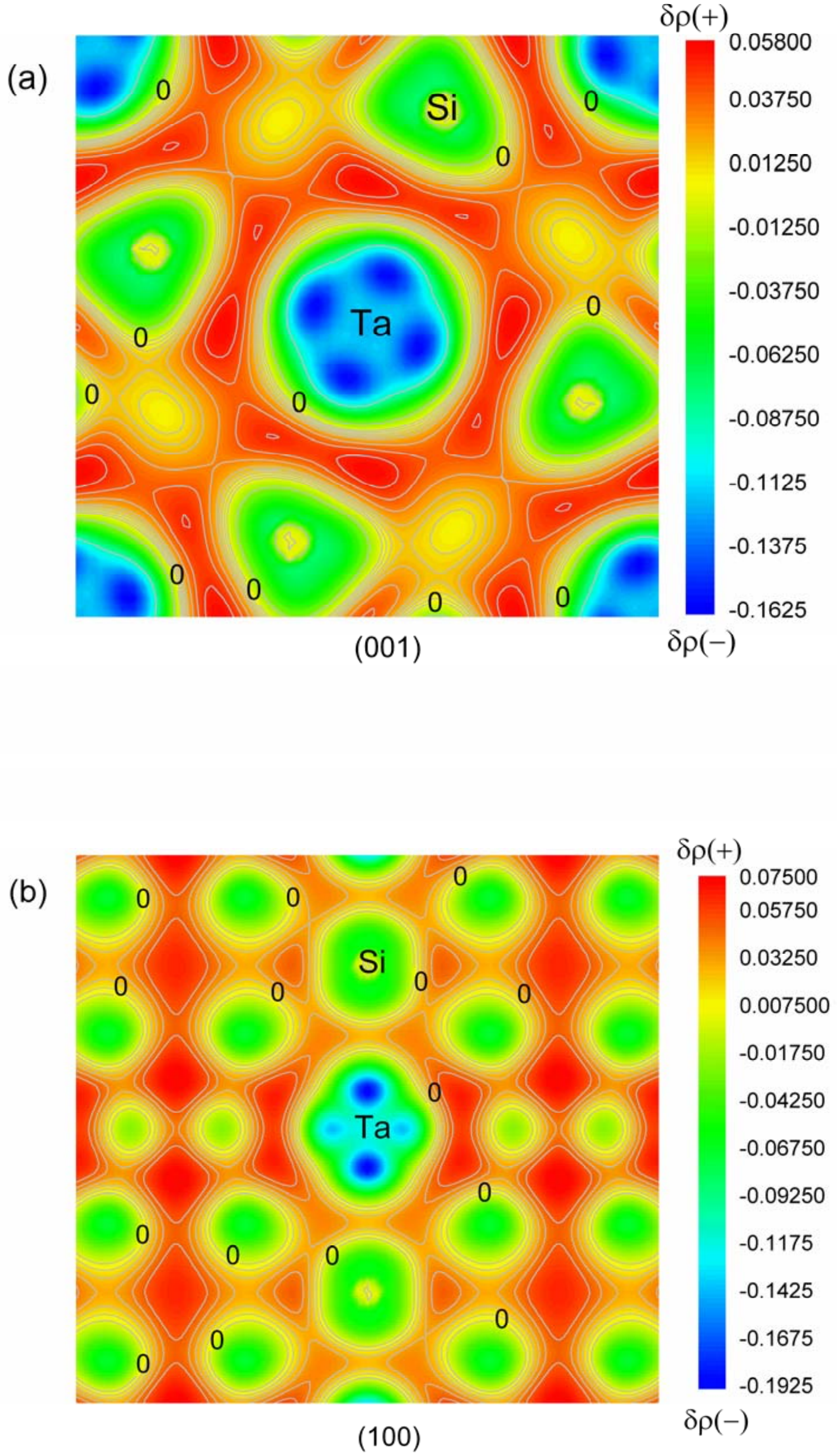

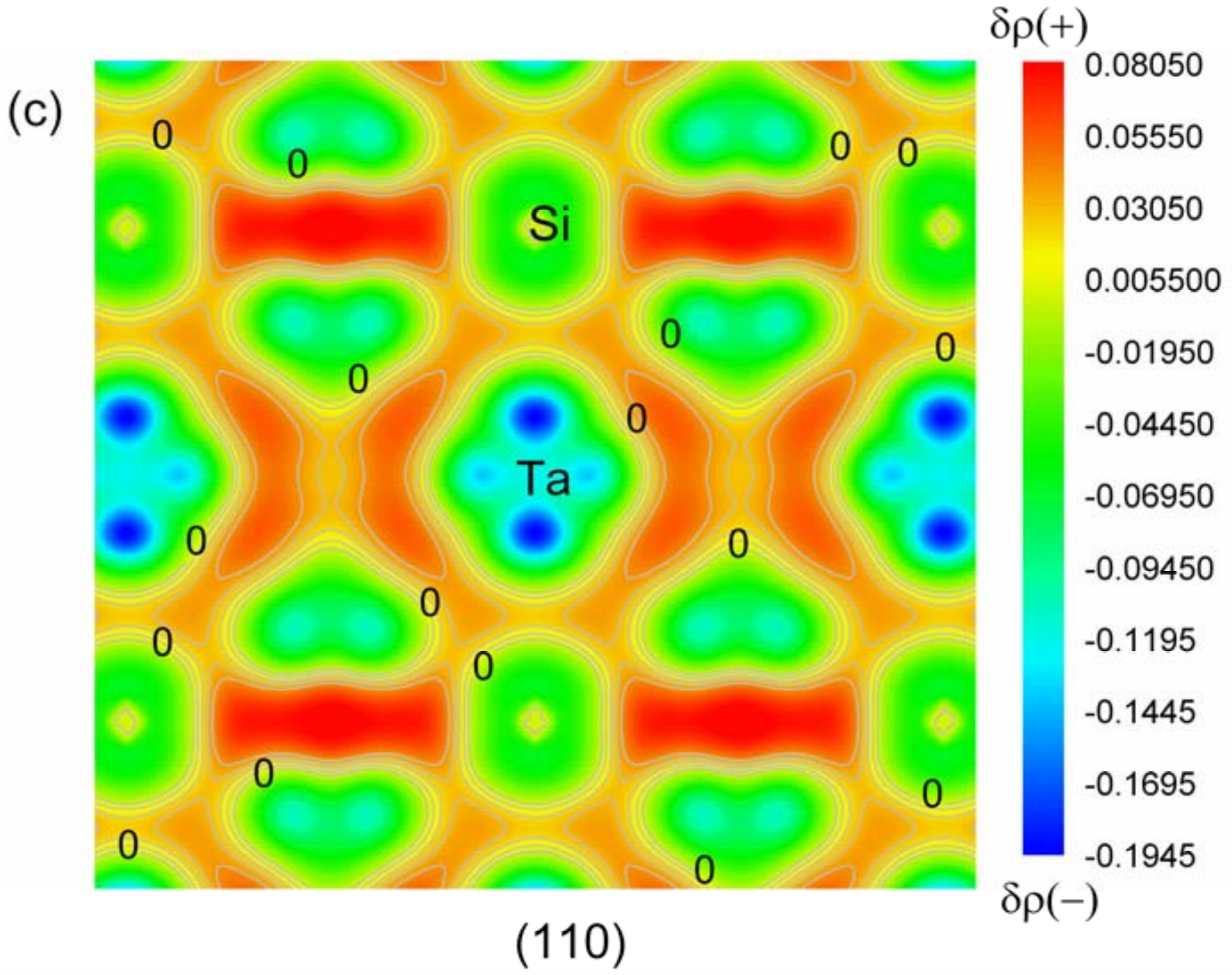

Figure 4

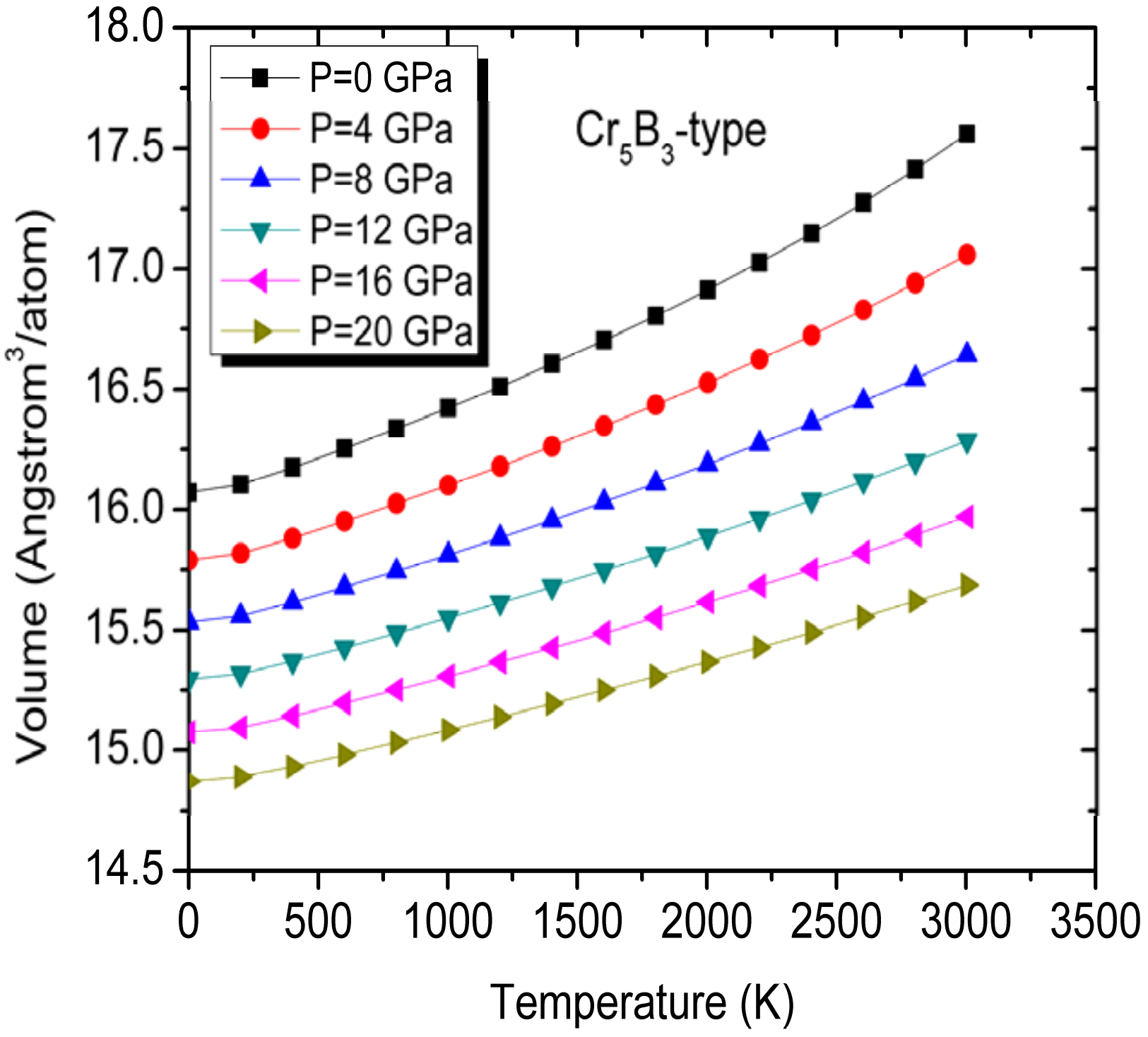

Figure 5

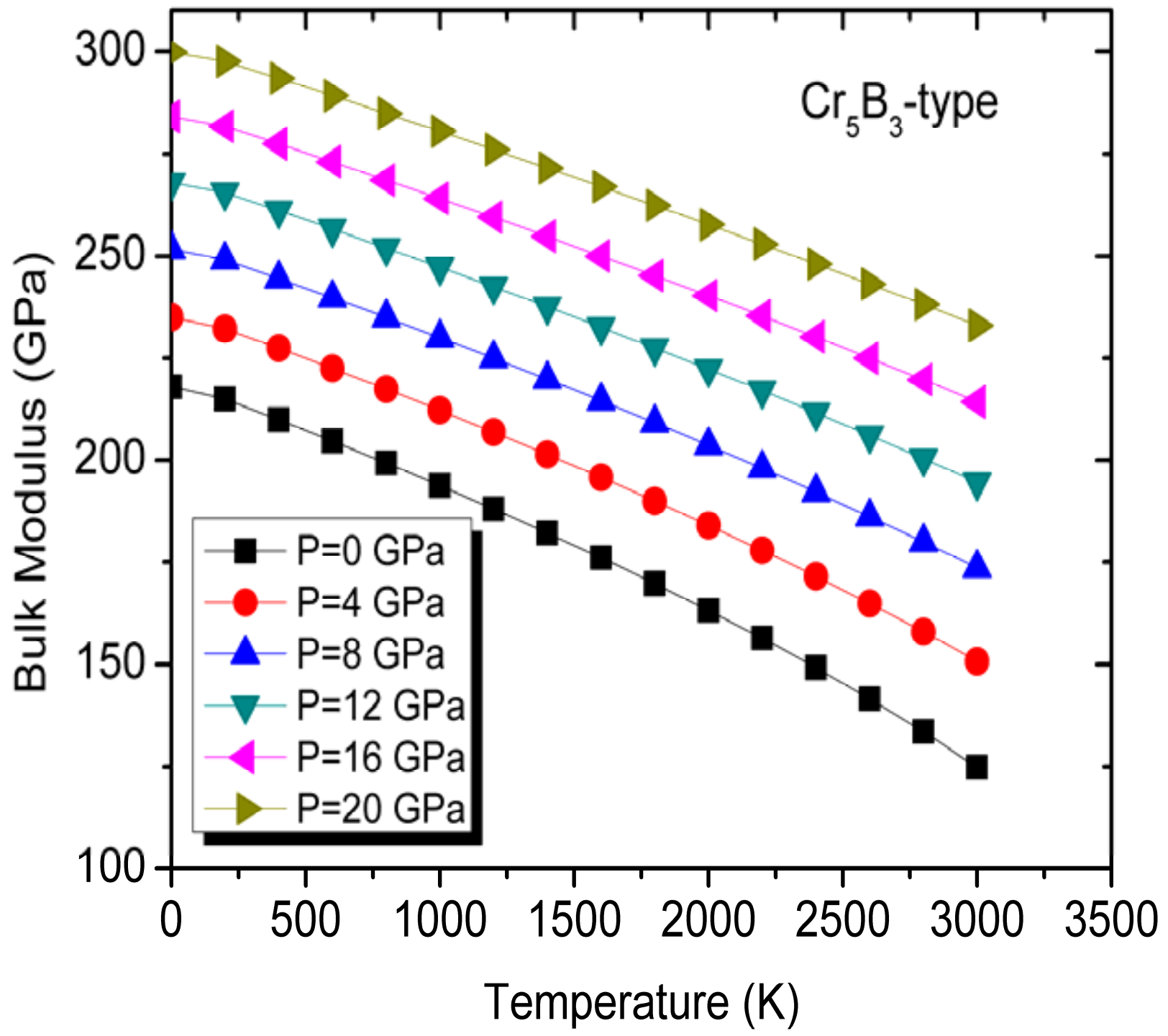

Figure 6

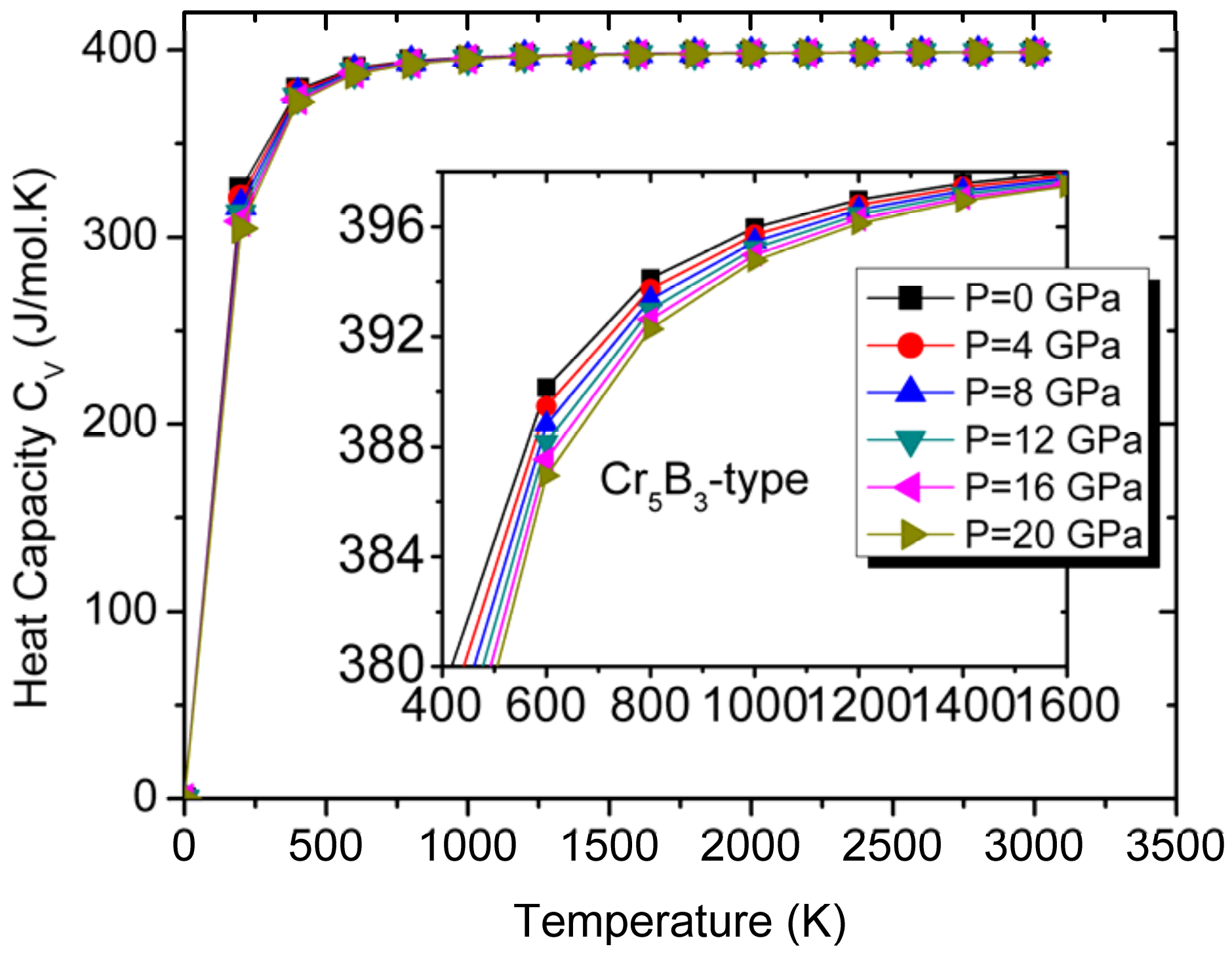

Figure 7

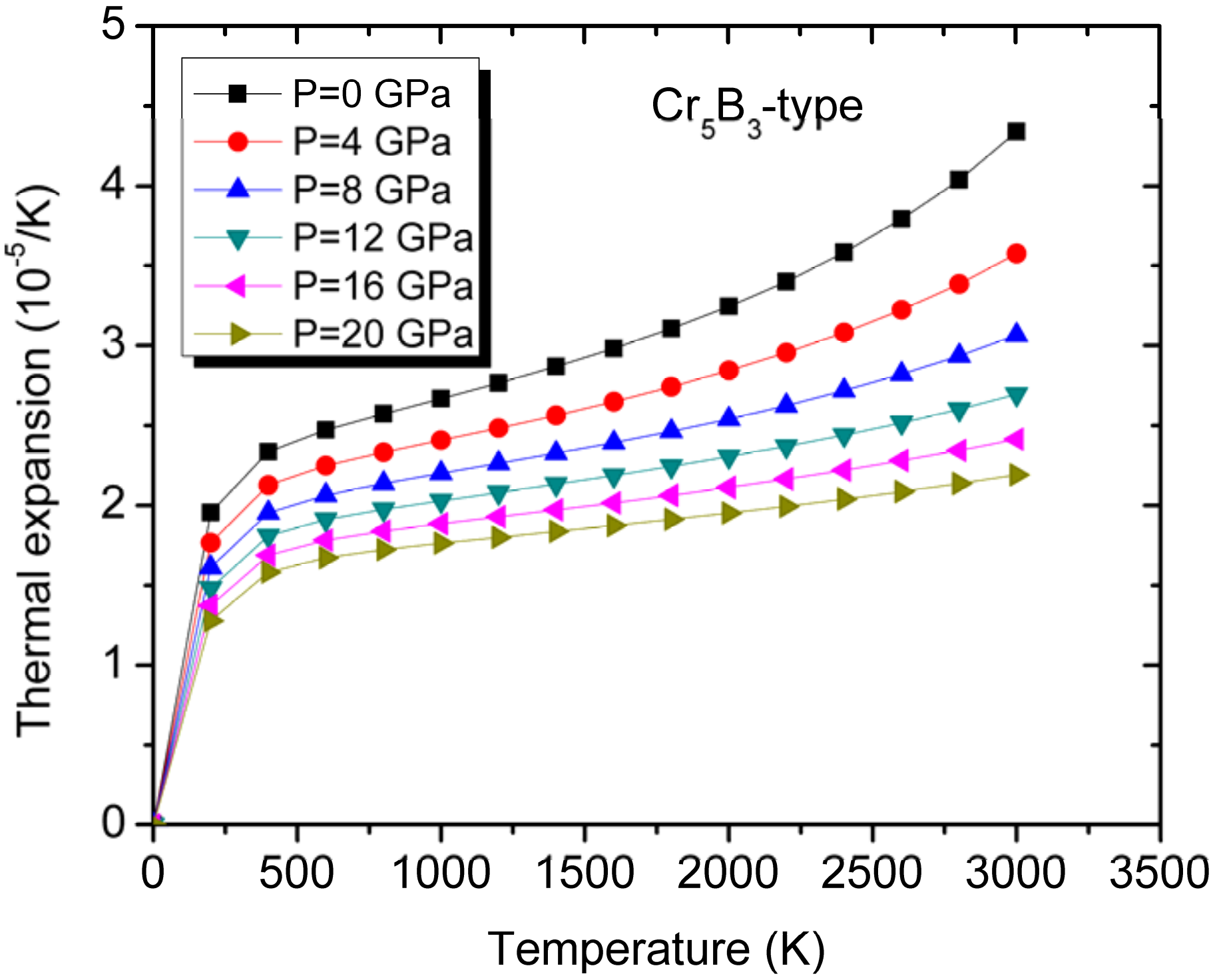

Figure 8

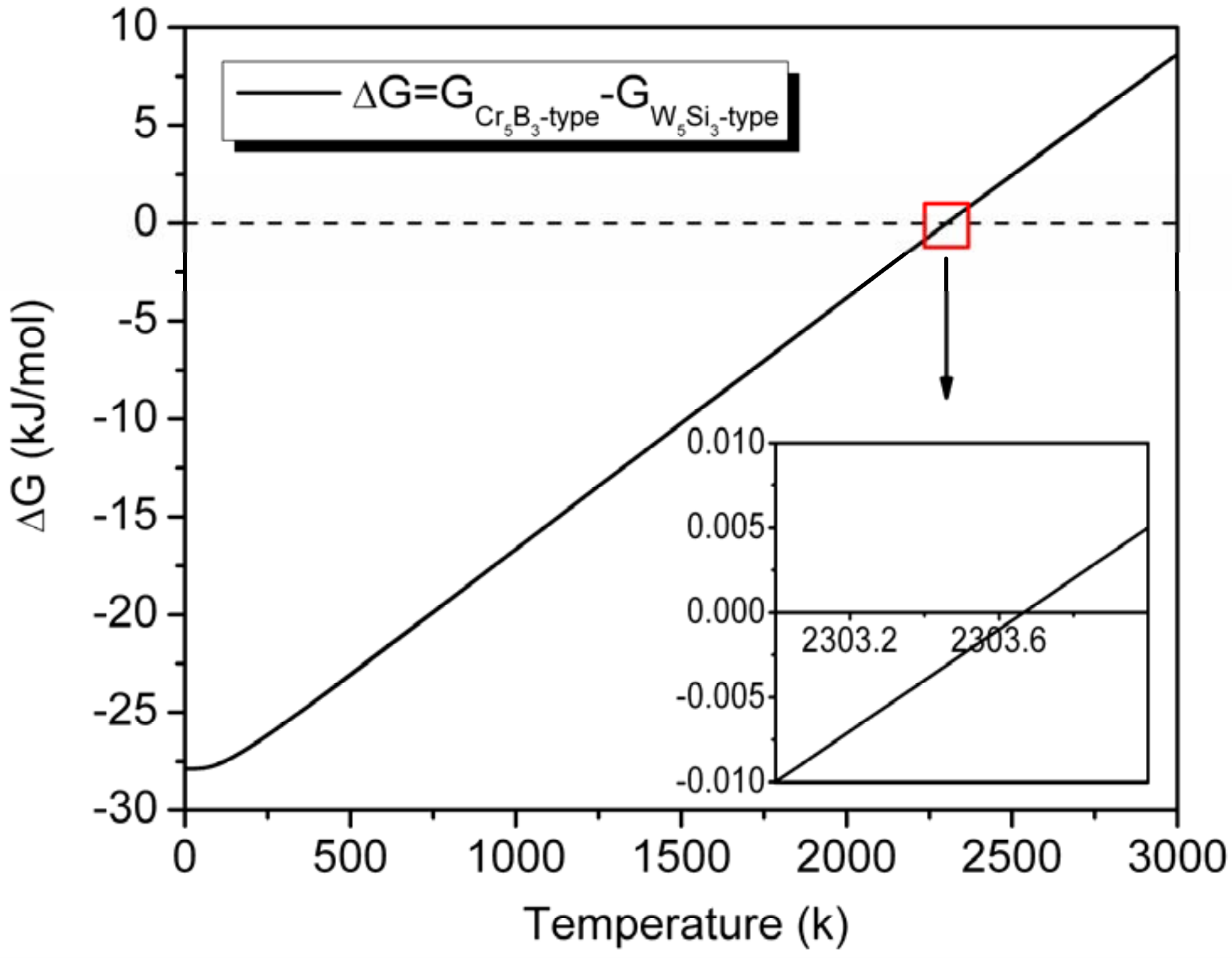

Figure 9